\documentclass[showpacs,eqsecnum,aps,prd,12pt]{revtex4}
\usepackage{graphicx}
\usepackage{amsmath}
\usepackage{amsfonts}
\usepackage{amssymb}
\usepackage{ifthen}
\usepackage{fancyheadings}

\newcommand{\ignore}[1]{}
\newcommand{\alternative}[2]{#2}

\renewcommand{\emph}{\textsl}

\newcommand{\latin}{\textit}

\renewcommand{\d}{\mathrm{d}}
\newcommand{\diff}[2]{
  \ifthenelse{\equal{#1}{}}%
{\frac{\mathrm{d}\hphantom{#2}}{\mathrm{d}#2}}%
{\frac{\mathrm{d}#1}{\mathrm{d}#2}}}
\newcommand{\ddiff}[2]{
  \ifthenelse{\equal{#1}{}}%
{\frac{\mathrm{d}^2\hphantom{#2}}{\mathrm{d}#2^2}}
{\frac{\mathrm{d}^2#1}{\mathrm{d}#2^2}}}
\newcommand{\pardiff}[2]{
	\ifthenelse{\equal{#1}{}}%
{\frac{\partial\phantom{#2}}{\partial#2}}
{\frac{\partial#1}{\partial#2}}
}

\renewcommand{\vec}[1]{\boldsymbol{#1}}

\newcommand{\vacrm}[3][]{\left\langle#2\right\rangle^{\mathrm{#3}}_{\mathrm{#1}}}
\newcommand{\vac}[3][]{\left\langle#2\right\rangle^{#3}_{\mathrm{#1}}}
\newcommand{\abs}[1]{\left|#1\right|}
\newcommand{\Green}[3][]{#2^{#3}_{\mathrm{#1}}}

\begin{document}
\pagenumbering{arabic}

\title{The Rotating Quantum Thermal Distribution}
\author{Gavin Duffy}
\email{Gavin.Duffy@ucd.ie}
\author{Adrian C. Ottewill}
\email{ottewill@relativity.ucd.ie}
\affiliation{Department of Mathematical Physics, University College Dublin,
Belfield, Dublin 4, Ireland.}

\begin{abstract}
We show that the rigidly rotating quantum thermal distribution on flat
space-time suffers from a global pathology which can be cured by introducing a
cylindrical mirror if and only if it has a radius smaller than that of the
speed-of-light cylinder. When this condition is met, we demonstrate numerically
that the renormalized expectation value of the energy-momentum stress tensor
corresponds to a rigidly rotating thermal bath up to a finite correction except
on the mirror where there are the usual Casimir divergences.
\end{abstract}


\maketitle

\section{Introduction}
In a recent investigation of the rotating quantum vacuum, Davies
\latin{et.~al.}~\cite{ar:Davies96} uncovered a remarkable issue involving the
speed-of-light cylinder of flat space-time. A particle detector rotating at the
same angular velocity as this vacuum fails to remain inert in unbounded
space-time and remains inert when the space-time is bounded by an infinite
cylindrical mirror only if the mirror has a radius less than that of the
speed-of-light cylinder. It is only in this case that the concept of a rotating
vacuum is unambiguously defined. In the present article, we attempt to extend
this discussion of rotation to the problem of defining a rigidly rotating
quantum thermal distribution. Whereas we might have expected that this
distribution would be pathologically only on and outside the speed-of-light
cylinder, in fact we find that it is pathological almost everywhere on the
unbounded space-time and likewise on the space-time bounded by a cylindrical
mirror except when the mirror has a radius less than that of the speed-of-light
cylinder.

This problem is closely related to the definition of a Hartle-Hawking vacuum on
Kerr space-time. In particular, on the asymptotically flat region of the
space-time for a black hole which is rotating arbitrarily slowly, the
contributions made to the field by the upgoing modes become negligible and when
these are discarded the anti-commutator function associated with the
Hartle-Hawking vacuum of~\cite{ar:Frolov89} coincides with the thermal
distribution considered here.

\section{The Distribution on Unbounded Space-Time}
We first introduce a cylindrical co-ordinate system $\{t_{+},R,\varphi_{+},z\}$
rigidly rotating at a fixed angular velocity $\Omega$; this is related to the
usual cylindrical Minkowski co-ordinate system $\{t,R,\varphi,z\}$ by the
transformation
\begin{equation}
t_{+}=t,\qquad
\varphi_{+}=\varphi-\Omega t.
\end{equation}
These co-ordinates are appropriate to the discussion of rigidly rotating
observers (RROs) in flat space-time. These are analogous to observers
co-rotating with the horizon in Kerr space-time. The scalar wave equation is
separable in these co-ordinates. Indeed the corresponding positive Klein-Gordon
norm modes,
\alternative{
\begin{equation}
\tilde{u}_{\tilde\omega km}=\frac{1}{\sqrt{8\pi^2}}
	e^{-i\tilde\omega t_{+}+im\varphi_{+}+ikz}
  J_m\left(\sqrt{(\tilde\omega+m\Omega)^2-k^2}R\right),
\label{eq:rotModes}
\end{equation}
}{
\begin{multline}
\tilde{u}_{\tilde\omega km}=\frac{1}{\sqrt{8\pi^2}}
	e^{-i\tilde\omega t_{+}+im\varphi_{+}+ikz}\times\\
  J_m\left(\sqrt{(\tilde\omega+m\Omega)^2-k^2}R\right),
\label{eq:rotModes}
\end{multline}
}
may be obtained from the standard modes of Minkowski space-time in cylindrical
co-ordinates,
\begin{equation}
u_{\omega km}=\frac{1}{\sqrt{8\pi^2}}e^{-i\omega t+im\varphi+ikz}
  J_m\left(\sqrt{\omega^2-k^2}R\right),
\label{eq:normalModes}
\end{equation}
by the identification
\begin{equation}
\tilde\omega=\omega-m\Omega.
\end{equation}
Since it is the norms not the frequencies of the RRO modes which determine the
commutation relations of the associated creation and annihilation operators,
the rotating vacuum naively coincides with the conventional vacuum of Minkowski
space-time.

The thermal distribution at inverse temperature $\beta$ and rigidly rotating at
angular velocity $\Omega$ will be described by the density operator
\begin{equation}
\hat{\rho}_{+}=e^{-\beta\hat{H}_{+}},
\label{eq:densityOperator}
\end{equation}
with the Hamiltonian
\begin{equation}
\hat{H}_{+}=i\pardiff{}{t_{+}}.
\label{eq:Hamiltonian}
\end{equation}
The thermal anti-commutator function $\Green[(1)]{G}{\beta}$ associated with
this distribution is therefore characterized by the
condition~\cite{bk:Birrell82}
\begin{equation}
\Green[(1)]{G}{\beta}(t_{+},\vec{x};t'_{+},\vec{x}')=
  \Green[(1)]{G}{\beta}(t_{+}+i\beta,\vec{x};t'_{+},\vec{x}').
\label{eq:KMS}
\end{equation}
It can be checked that at least formally this condition is satisfied by the
anti-commutator function defined by the mode sum
\alternative{
\begin{multline}
\Green[(1)]{G}{\beta}(x,x')=
	\frac{1}{8\pi^2}\sum_{m=-\infty}^{\infty}\int_{-\infty}^\infty\d{\omega}
	\int_{0}^{\omega}\d{k}\,
	\coth\left(\frac{\beta\tilde\omega}{2}\right)
	e^{im(\varphi-\varphi')+ik(z-z')}\Re\left[e^{-i\omega(t-t')}\right]\times\\
	J_m\left(\sqrt{\omega^2-k^2}R\right)J_m\left(\sqrt{\omega^2-k^2}R'\right).
\label{eq:twoPointPathological}
\end{multline}
}{
\begin{multline}
\Green[(1)]{G}{\beta}(x,x')=
	\frac{1}{8\pi^2}\sum_{m=-\infty}^{\infty}\int_{-\infty}^\infty\d{\omega}
	\int_{0}^{\omega}\d{k}\,
	\coth\left(\frac{\beta\tilde\omega}{2}\right)\times\\
	e^{im(\varphi-\varphi')+ik(z-z')}\Re\left[e^{-i\omega(t-t')}\right]\times\\
	J_m\left(\sqrt{\omega^2-k^2}R\right)J_m\left(\sqrt{\omega^2-k^2}R'\right).
\label{eq:twoPointPathological}
\end{multline}
}
As the energy $\tilde\omega$ as measured by $\hat{H}_{+}$ tends towards zero,
the density of states factor $\coth(\beta\tilde\omega/2)$ becomes infinite
although the modes themselves remain non-zero whenever $m\ne0$. These modes
clearly make divergent contributions to the mode
sum~(\ref{eq:twoPointPathological}) except when either $R$ or $R'$ is zero. The
anti-commutator function $\Green[(1)]{G}{\beta}$ is thus pathological except
when at least one of the two points is on the $z$-axis.
In~\cite{ar:Ottewill00}, a similar pathology was noted in the mode
sum for the anti-commutator function of the Hartle-Hawking state considered
in~\cite{ar:Frolov89}.

\section{The Distribution within an Infinite Cylinder}
We now introduce a cylindrical mirror of arbitrary radius $R_{0}$. For brevity,
we only treat the case of a field which satisfies Dirichlet conditions on this
cylinder. Introducing a non-dimensional radial variable $\bar{R}=R/R_0$, a
complete set of orthonormal solutions to the field equation subject to the
boundary conditions is
\alternative{
\begin{equation}
u_{kmn}=
	\frac{1}{2\pi R_{0}\sqrt{|\omega_{kmn}|}\abs{J_{m+1}\left(\xi_{mn}\right)}}
	e^{-i\omega_{kmn} t+im\varphi+ikz}J_m\left(\xi_{mn}\bar R\right),
\label{eq:Minmodes}
\end{equation}
}{
\begin{multline}
u_{kmn}=
	\frac{1}{2\pi R_{0}\sqrt{|\omega_{kmn}|}\abs{J_{m+1}\left(\xi_{mn}\right)}}
	\times\\
	e^{-i\omega_{kmn} t+im\varphi+ikz}J_m\left(\xi_{mn}\bar R\right),
\label{eq:Minmodes}
\end{multline}
}
where
\begin{equation}
\omega_{kmn}=\pm\sqrt{\frac{\xi_{mn}^2}{R_{0}^2}+k^2}
\label{eq:omegaDiscrete}
\end{equation}
and $\xi_{mn}$ denotes the $n^{\mathrm{th}}$ positive zero of $J_m$. The
normalization factor has been calculated by making use of an identity for the
Bessel functions~\cite[see page 765]{bk:Morse53}. The modes which have positive
norm are precisely those which have positive frequency $\omega$. The vacuum
state associated with the field when it is expanded in terms of this set of
modes has a anti-commutator function which is given by the mode sum
\alternative{
\begin{multline}
\Green[(1)]{G}{}(x,x')=\sum_{m=-\infty}^\infty\sum_{n=1}^{\infty}
	\int_{-\infty}^\infty\d{k}\,e^{im(\varphi-\varphi')+ik(z-z')}\times\\
	\frac{J_m\left(\xi_{mn}\bar R\right)J_m\left(\xi_{mn}\bar R'\right)}%
{4\pi^2R_{0}^2\omega_{kmn}J_{m+1}^2\left(\xi_{mn}\right)}
	2\Re\left[e^{-i\omega_{kmn}(t-t')}\right].
\label{eq:vacuumTwoPoint}
\end{multline}
}{
\begin{multline}
\Green[(1)]{G}{}(x,x')=\sum_{m=-\infty}^\infty\sum_{n=1}^{\infty}
	\int_{-\infty}^\infty\d{k}\,e^{im(\varphi-\varphi')+ik(z-z')}\times\\
	\frac{J_m\left(\xi_{mn}\bar R\right)J_m\left(\xi_{mn}\bar R'\right)}%
{4\pi^2R_{0}^2\omega_{kmn}J_{m+1}^2\left(\xi_{mn}\right)}
	2\Re\left[e^{-i\omega_{kmn}(t-t')}\right].
\label{eq:vacuumTwoPoint}
\end{multline}
}
The anti-commutator function
\alternative{
\begin{multline}
\Green[(1)]{G}{\beta}(x,x')=
	\sum_{m=-\infty}^\infty\sum_{n=1}^{\infty}
	\int_{-\infty}^\infty\d{k}\,
	\coth\left(\frac{\beta\tilde\omega_{kmn}}{2}\right)
	e^{im(\varphi-\varphi')+ik(z-z')}\times\\
	\frac{J_m\left(\xi_{mn}\bar R\right)J_m\left(\xi_{mn}\bar R'\right)}%
{4\pi^2R_{0}^2\omega_{kmn}J_{m+1}^2\left(\xi_{mn}\right)}
	2\Re\left[e^{-i\omega_{kmn}(t-t')}\right]
\label{eq:thermalTwoPoint}
\end{multline}
}{
\begin{multline}
\Green[(1)]{G}{\beta}(x,x')=
	\sum_{m=-\infty}^\infty\sum_{n=1}^{\infty}
	\int_{-\infty}^\infty\d{k}\,
	\coth\left(\frac{\beta\tilde\omega_{kmn}}{2}\right)\times\\
	e^{im(\varphi-\varphi')+ik(z-z')}\times\\
	\frac{J_m\left(\xi_{mn}\bar R\right)J_m\left(\xi_{mn}\bar R'\right)}%
{4\pi^2R_{0}^2\omega_{kmn}J_{m+1}^2\left(\xi_{mn}\right)}
	2\Re\left[e^{-i\omega_{kmn}(t-t')}\right]
\label{eq:thermalTwoPoint}
\end{multline}
}
is associated with a thermal distribution described by the density
operator~(\ref{eq:densityOperator}) with Hamiltonian~(\ref{eq:Hamiltonian}).
This mode sum suffers from a similar pathology to that on the unbounded
space-time unless there are no positive frequency modes for which
$\tilde\omega_{kmn}$ is zero. It is a well known property of the zeros of $J_m$
that $\xi_{m1}>|m|$~\cite[see \S15.3]{bk:Watson22} and so we see
from~(\ref{eq:omegaDiscrete}) that if $R_0<\Omega^{-1}$, no such modes exist.
On the other hand, the asymptotic behaviour of this first zero
is~\cite[see page xviii]{bk:Olver60}
\begin{equation}
\xi_{m1}\sim m+1.85575m^{1/3},\qquad(m\to\infty),
\end{equation}
from which we see that if $R_0>\Omega^{-1}$, there are modes of this type for
all sufficiently large $m$. It follows that~(\ref{eq:thermalTwoPoint}) is well
behaved if and only if the mirror lies within the speed-of-light cylinder.

\section{The Measurements of an RRO}
\begin{figure}
\centering
\includegraphics*{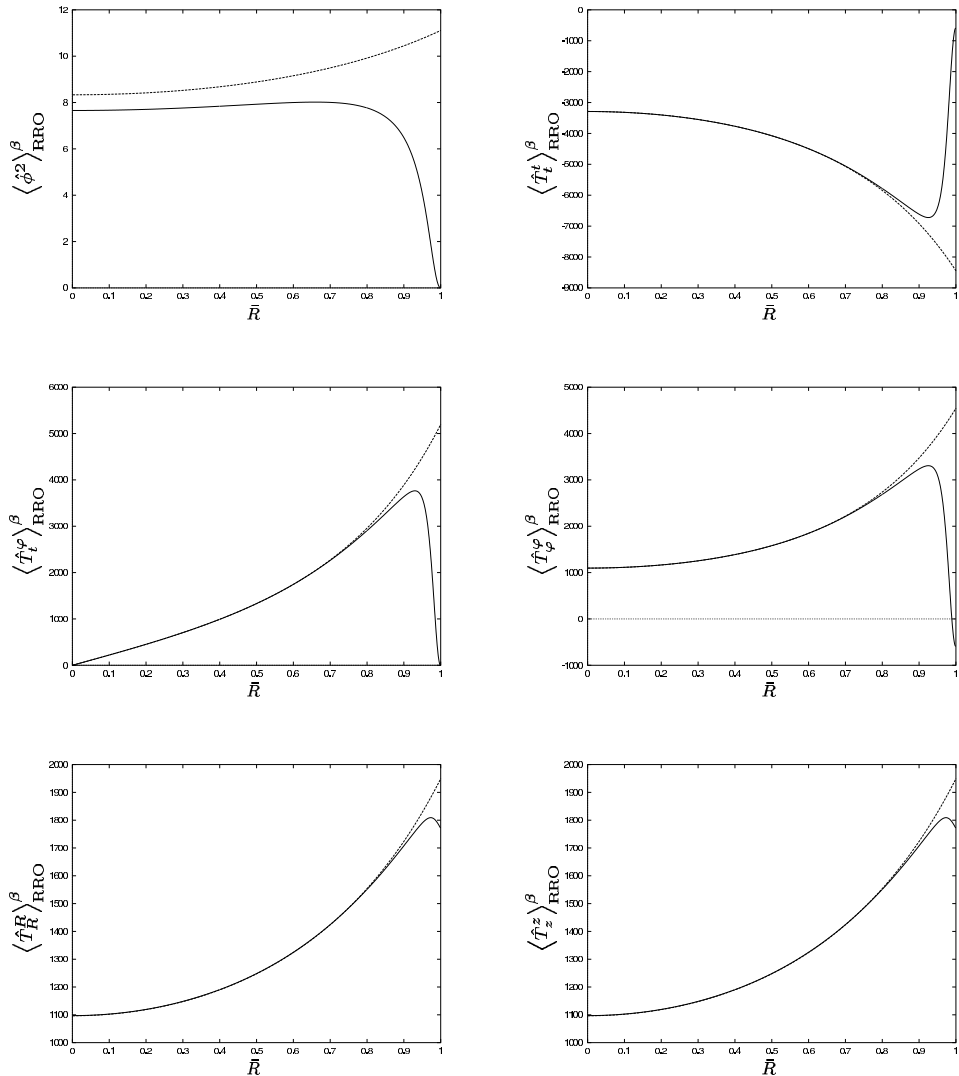}
\caption{The graphs are given in units in which $R_0$, the radius of the
cylinder, is unity. The temperature is $T=10/R_0$ and the angular velocity is
$\Omega=0.5/R_0$. The dashed line is a plot of the value for a rigidly rotating
thermal distribution~(\ref{eq:phirot}--\ref{eq:Tzzrot}).}
\label{fig:ThermalExpct}
\end{figure}

When the mirror lies inside the speed-of-light cylinder, static observers and
RROs both make measurements with respect to the vacuum state whose
anti-commutator function is given in~(\ref{eq:vacuumTwoPoint}). These
measurements can be calculated from
\alternative{
\begin{multline}
\Green[(1)]{G}{\beta}(x,x')-\Green[(1)]{G}{}(x,x')=
	\sum_{m=-\infty}^\infty\sum_{n=1}^{\infty}
	\int_{-\infty}^\infty\d{k}\,\frac{1}{e^{\beta\tilde\omega_{kmn}}-1}
	e^{im(\varphi-\varphi')+ik(z-z')}\times\\
	\frac{J_m\left(\xi_{mn}\bar R\right)J_m\left(\xi_{mn}\bar R'\right)}%
{4\pi^2R_{0}^2\omega_{kmn}J_{m+1}^2\left(\xi_{mn}\right)}
	2\Re\left[e^{-i\omega_{kmn}(t-t')}\right].
\label{eq:Ddiff}
\end{multline}
}{
\begin{multline}
\Green[(1)]{G}{\beta}(x,x')-\Green[(1)]{G}{}(x,x')=\\
	\sum_{m=-\infty}^\infty\sum_{n=1}^{\infty}
	\int_{-\infty}^\infty\d{k}\,\frac{1}{e^{\beta\tilde\omega_{kmn}}-1}
	e^{im(\varphi-\varphi')+ik(z-z')}\times\\
	\frac{J_m\left(\xi_{mn}\bar R\right)J_m\left(\xi_{mn}\bar R'\right)}%
{4\pi^2R_{0}^2\omega_{kmn}J_{m+1}^2\left(\xi_{mn}\right)}
	2\Re\left[e^{-i\omega_{kmn}(t-t')}\right].
\label{eq:Ddiff}
\end{multline}
}
We can derive from this a set of expressions for the non-zero components of the
energy-momentum stress tensor corresponding to the the conformally invariant
field, the details of which can be found in~\cite{th:Duffy02}. The mode by mode
cancellation of the high frequency divergences which afflict both
anti-commutator functions in the coincident limit makes these expressions
amenable to numerical analysis and the results are shown in
figure~\ref{fig:ThermalExpct}. They are compared with the Planckian forms
corresponding to a rigidly rotating thermal distribution at temperature $T$
which are
\begin{align}
\vacrm[Planck]{\phi^2}{\beta}&=\frac{(\gamma T)^2}{12},\label{eq:phirot}\\
\vacrm[Planck]{ T_t^t}{\beta}&=-\frac{\pi^2}{90}\left(3+v^2\right)\gamma^2
	(\gamma T)^4,\label{eq:Tttrot}\\ 
\vacrm[Planck]{ T_t^\varphi}{\beta}&=\frac{4\pi^2}{90}v\gamma^2(\gamma T)^4,
	\label{eq:Ttphirot}\\
\vacrm[Planck]{ T_\varphi^\varphi}{\beta}&=\frac{\pi^2}{90}
	\left(1+3v^2\right)\gamma^2(\gamma T)^4,\label{eq:Tphiphirot}\\
\vacrm[Planck]{ T_R^R}{\beta}&=\frac{\pi^2}{90}(\gamma T)^4,
\label{eq:TRRrot}\\
\vacrm[Planck]{ T_z^z}{\beta}&=\frac{\pi^2}{90}(\gamma T)^4,\label{eq:Tzzrot}
\end{align}
where $v$ and $\gamma$ are given by
\begin{equation}
v=R\Omega,\qquad \gamma=\frac{1}{\sqrt{1-v^2}}
\end{equation}
and are the speed and Lorentz factor of an RRO at the appropriate space-time
point. We find that they are in close agreement everywhere except, as expected,
close to the mirror.

\section{Renormalized Expectation Values}
\begin{figure}
\centering
\includegraphics*{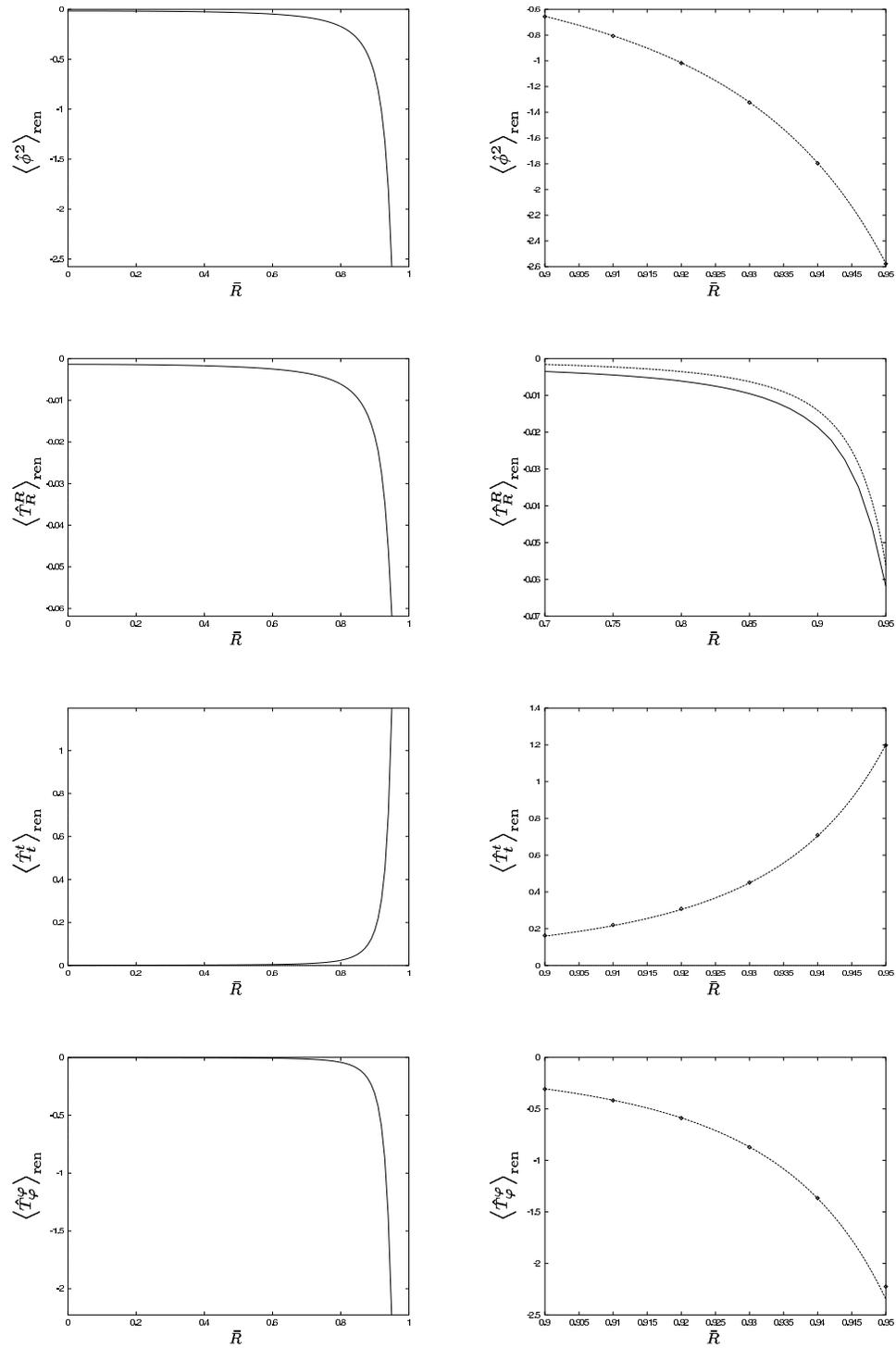}
\caption{The graphs are given in units in which $R_0$, the radius of the
cylinder, is unity. In the right hand graphs, the dashed line gives the
analytically calculated Casimir
divergence~(\ref{eq:phiCas}--\ref{eq:TphiphiCas}) while the solid line in the
second graph and the points in the others give the numerically calculated
values.}
\label{fig:RenExpct}
\end{figure}

A renormalized expectation value differs from that an RRO measures by a term
due to polarization of the vacuum by the mirror. This term can be calculated by
making use of the relationship between that Feynman propagator and the
Euclidean Green function; on the Euclidean section of the manifold, the
analysis becomes essentially identical to that of a uniformly accelerating
infinite flat mirror on the Euclidean section of the Rindler manifold and we
can proceed along the lines of~\cite{ar:Candelas77a}. We find that the
Euclidean Green function which vanishes on the mirror is
\alternative{
\begin{multline}
\Green[E]{G}{}(x,x')=\int_{-\infty}^{\infty}
	\frac{\d{\omega}}{2\pi}e^{-i\omega(\tau-\tau')}
	\int_{-\infty}^{\infty}\frac{\d{k}}{2\pi}e^{ik(z-z')}
	\sum_{m=-\infty}^{\infty}\frac{e^{im(\varphi-\varphi')}}{2\pi}\times\\
	I_m(\alpha R_<)\left\{K_m(\alpha R_>)-
		I_m(\alpha R_>)\frac{K_m(\alpha R_{0})}{I_m(\alpha R_{0})}\right\},
\label{eq:EuclideanGreen}
\end{multline}
}{
\begin{multline}
\Green[E]{G}{}(x,x')=\\
	\int_{-\infty}^{\infty}
	\frac{\d{\omega}}{2\pi}e^{-i\omega(\tau-\tau')}
	\int_{-\infty}^{\infty}\frac{\d{k}}{2\pi}e^{ik(z-z')}
	\sum_{m=-\infty}^{\infty}\frac{e^{im(\varphi-\varphi')}}{2\pi}\times\\
	I_m(\alpha R_<)\left\{K_m(\alpha R_>)-
		I_m(\alpha R_>)\frac{K_m(\alpha R_{0})}{I_m(\alpha R_{0})}\right\},
\label{eq:EuclideanGreen}
\end{multline}
}
where $R_<=\min\{R,R'\}$, $R_>=\max\{R,R'\}$ and $t=i\tau$. The second term in
the braces is absent in the case of the Euclidean Green function on the
unbounded manifold and so this is the term that remains after renormalization.
Now, closing the points and making a transformation to polar variables $\alpha$
and $\gamma$ defined by
\begin{equation}
kR_0=\alpha\sin\gamma,\qquad\omega R_0=\alpha\cos\gamma,
\end{equation}
we find that we can perform the integral over $\gamma$ to obtain
\begin{equation}
\vac[ren]{\hat\phi^2}{}=-\frac{1}{4\pi^2R_{0}^2}\sum_{m=-\infty}^{\infty}
	\int_{0}^{\infty}\d{\alpha}\,\alpha I_m^2(\alpha\bar R)
	\frac{K_m(\alpha)}{I_m(\alpha)}.
\label{eq:phicyl}
\end{equation}
A similar thing can be done for the components of the energy-momentum stress
tensor and the resulting expressions together with~(\ref{eq:phicyl}) are
amenable to numerical analysis. Once again, the details can be found
in~\cite{th:Duffy02}. The results are presented in figure~\ref{fig:RenExpct}.
They are compared with the Casimir divergence close to the mirror which can be
calculated by an asymptotic analysis following~\cite{ar:Candelas77a}. The
relevant expressions are
\begin{alignat}{2}
\vac[ren]{\hat\phi^2}{}&\sim
	-\frac{1}{16\pi^2R_{0}^2(1-\bar R)^2}\left[1+\frac{1-\bar R}{3}\right],&
	\qquad(\bar{R}&\to1).\label{eq:phiCas}\\
\vac[ren]{\hat T^R_R}{}&\sim-\frac{1}{720\pi^2R_{0}^4(1-\bar R)^2},&
	\qquad(\bar{R}&\to1),\label{eq:TRRCas}\\
\vac[ren]{\hat T^t_t}{}&\sim\frac{1}{720\pi^2R_{0}^4(1-\bar R)^3}
	\left[1+\frac{19(1-\bar R)}{14}\right],&
	\qquad(\bar{R}&\to1),\label{eq:TttCas}\\
\vacrm[ren]{\hat T^\varphi_\varphi}{}&\sim-\frac{1}{360\pi^2R_{0}^4(1-\bar R)^3}
	\left[1+\frac{6(1-\bar R)}{7}\right],&
	\qquad(\bar{R}&\to1).\label{eq:TphiphiCas}
\end{alignat}
%\begin{alignat}{2}
%\vac[ren]{\hat\phi^2}{}&\sim
%	-\frac{1}{16\pi^2(R_{0}-R)^2}\left[1+\frac{(R_{0}-R)}{3R_{0}}\right],&
%\qquad(R&\to R_{0}),\label{eq:phiCas}\\
%\vac[ren]{\hat T^R_R}{}&\sim-\frac{1}{720\pi^2R^2_{0}(R_{0}-R)^2},&
%	\qquad(R&\to R_{0}),\label{eq:TRRCas}\\
%\vac[ren]{\hat T^t_t}{}&\sim\frac{1}{720\pi^2R_{0}(R_{0}-R)^3}
%	\left[1+\frac{19(R_{0}-R)}{14R_{0}}\right],&
%	\qquad(R&\to R_{0}),\label{eq:TttCas}\\
%\vacrm{\hat T^\varphi_\varphi}{ren}&\sim-\frac{1}{360\pi^2R_{0}(R_{0}-R)^3}
%	\left[1+\frac{6(R_{0}-R)}{7R_{0}}\right],&
%	\qquad(R&\to R_{0}).\label{eq:TphiphiCas}
%\end{alignat}
and are in agreement with the general expressions of~\cite{ar:Deutsch79}. We
have used these to calculate renormalized expectation values in the thermal
distribution when $\Omega=0$ and checked that they are well approximated by the
general expressions given in~\cite{ar:Kennedy80}.

\section{Conclusion}
We found that the anti-commutator function associated with the rigidly rotating
thermal distribution on unbounded Minkowski space-time is pathological almost
everywhere. The pathology is caused by the existence of non-zero modes which
have zero energy as measured by the Hamiltonian $\hat{H}_{+}$ relevant to RROs.
In~\cite{ar:Ottewill00}, a similar pathology was noted in the anti-commutator
function of the Hartle-Hawking state considered in~\cite{ar:Frolov89}. In this
case $\hat{H}_{+}$ is the Hamiltonian relevant to observers rigidly rotating
with the horizon. The corresponding modes are thus at the critical point of
superradiant scattering. When Minkowski space-time is bounded by an infinite
cylinder of radius larger than the speed-of-light cylinder we found that the
anti-commutator function is once again pathological almost everywhere because
of the existence of these modes for all sufficiently high $m$. In a future
article we will show that when the Kerr black hole is enclosed within a mirror
of constant Boyer-Lindquist radius larger than the minimum radius of the
speed-of-light surface, for all sufficiently high $m$ there are complex
frequency modes whose real parts lie in the regime we associate with
superradiance in the absence of the mirror. This set of modes has the critical
point of superradiant scattering as an accumulation point.

%\bibliography{%
%/home/duffy/tex/bibliography/papers,%
%/home/duffy/tex/bibliography/books,%
%/home/duffy/tex/bibliography/refs,%
%/home/duffy/tex/bibliography/refs_books%
%}
\end{document}